# Scale-Invariant Dissipationless Chiral Transport in Magnetic Topological Insulators beyond the Two-Dimensional Limit

Xufeng Kou,[1,†] Shih-Ting Guo,[2,†] Yabin Fan,[1,†] Lei Pan,[1] Murong Lang,[1] Ying Jiang,[3] Qiming Shao,[1] Tianxiao Nie,[1] Koichi Murata,[1] Jianshi Tang,[1] Yong Wang,[3] Liang He,[1] Ting-Kuo Lee,[2] Wei-Li Lee,[2] and Kang L. Wang [1]

[1]*Department of Electrical Engineering, University of California, Los Angeles, California 90095, USA*

[2]*Institute of Physics, Academia Sinica, Taipei 11529, Taiwan*

[3]*Center for Electron Microscopy and State Key Laboratory of Silicon Materials, Department of Materials Science and Engineering, Zhejiang University, Hangzhou 310027, China*

†These authors contributed equally to this work.

To whom correspondence should be addressed. E-mail: wlee@phys.sinica.edu.tw; wang@seas.ucla.edu

# Abstract

We investigate the quantum anomalous Hall Effect (QAHE) and related chiral transport in the millimeter-size $(Cr_{0.12}Bi_{0.26}Sb_{0.62})_2Te_3$ films. With high sample quality and robust magnetism at low temperatures, the quantized Hall conductance of $e^2/h$ is found to persist even when the film thickness is beyond the two-dimensional (2D) hybridization limit. Meanwhile, the Chern insulator-featured chiral edge conduction is manifested by the non-local transport measurements. In contrast to the 2D hybridized thin film, an additional weakly field-dependent longitudinal resistance is observed in the 10 quintuple-layer film, suggesting the influence of the film thickness on the dissipative edge channel in the QAHE regime. The extension of QAHE into the three-dimensional thickness region addresses the universality of this quantum transport phenomenon and motivates the exploration of new QAHE phases with tunable Chern numbers. In addition, the observation of the scale-invariant dissipationless chiral propagation on a macroscopic scale makes a major stride towards ideal low-power interconnect applications.

The discovery of topological insulator (TI) has greatly broadened the landscape of condensed matter physics [1-3]. Owing to a nontrivial band topology and strong spin-orbit coupling (SOC) in a three-dimensional (3D) TI material, gapless Dirac surface states protected by time-reversal-symmetry (TRS) are formed, and the surface conductions exhibit unusual spin-momentum locking feature [4-6]. In two-dimensional (2D) TIs, band inversion gives rise to the counter-propagating helical edge channels with opposite spins, and elastic back-scatterings from non-magnetic impurities are suppressed [7,8]. Accordingly, the resulting quantum spin Hall effect (QSHE) leads to a quantized longitudinal conductance of $2e^2/h$ in the absence of a magnetic field. However, since the dissipationless helical edge states are vulnerable to magnetic impurities and band potential fluctuations, TI materials with low defect density and high carrier mobility are crucial for the QSHE phase [9]. To date, the QSHE has only been experimentally observed in the 2D HgTe/CdTe [10,11] and InAs/GaSb quantum wells [12,13].

Alternatively, given the close connection between intrinsic anomalous Hall effect (AHE) and quantum Hall effect (QHE) in terms of the band topology [14-16], it is expected that a 2D ferromagnetic (FM) insulator with a non-zero first Chern number ($C_1$) would give rise to the quantized anomalous Hall effect (QAHE) [14]. In such Chern insulators, the chiral edge states are formed due to the TRS-breaking, and the spontaneous magnetization also localizes the dissipative states [17-19]. Among all possible candidates [20-24], it was proposed that by adding appropriate exchange splitting into the QSHE system, one set of the spin sub-bands would remain in the inversion regime while the other became topologically trivial, therefore driving the 2D magnetic TI system into a QAHE insulator [25-27]. Moreover, It was found that in magnetic tetradymite-type TI materials, robust out-of-plane magnetization could be developed directly from the large van Vleck spin susceptibility in the host TI materials without the mediation of itinerant carriers [27-29]. By manipulating the Fermi level position and the magnetic doping, QAHE in the 2D regime was recently observed in a five quintuple-layer (QL) $Cr_{0.15}(Bi_{0.1}Sb_{0.9})_{1.85}Te_3$ film, where a plateau of Hall conductance $\sigma_{xy}$ of $e^2/h$ and a vanishing longitudinal conductance $\sigma_{xx}$ were observed at 30 mK [30]. More generally, it has been suggested that the quantized Hall conductance could also be derived from the gapped top and bottom surfaces in 3D magnetic TI systems [7,31,32]. Furthermore, if the exchange field strength and film thickness were properly adjusted so that higher sub-bands would get involved in the band topology transition, and QAHE with a tunable Chern number could, in principle, be realized [33,34]. Nevertheless, the increased bulk conduction in the thicker films is detrimental and can obscure the observation of QAHE beyond the 2D hybridization thickness (> 6 QL) [30]. As a result, the universality of the QAHE phase and its related quantum transport phenomena in the 3D regime still remain unexplored. Here, we report the observation of QAHE in the $(Cr_{0.12}Bi_{0.26}Sb_{0.62})_2Te_3$ samples with the film thickness up to 10 QL. Given the chiral nature of the edge modes, we show that the quantization of the Hall conductance ($e^2/h$) persists in the device with millimeter-scale sizes. In contrast to the previous work [30], a non-zero longitudinal resistance is detected in our thick 10 QL sample and is found to be insensitive to external magnetic fields, indicating the possible presence of additional non-chiral side surface propagation modes. The corresponding non-local transport measurements further confirm the chiral property of a dissipationless QAHE state. This scale-invariant quantized chiral transport manifests the universality of the QAHE in both 2D and 3D thicknesses, and it may open up a practical route to construct new Chern insulators with higher Hall conductance plateaus [33-38].

To prepare the magnetic TI materials with pronounced FM orders and insulating bulk states, single-crystalline Cr-doped $(Bi_xSb_{1-x})_2Te_3$ films are grown by molecular beam epitaxy (MBE). Both the Cr doping level (12%) and the Bi/Sb ratio (0.3/0.7) are optimized so that the Fermi level positions of the as-grown samples are already close to the charge neutral point [29,39]. The growth is monitored by the reflection high-energy electron diffraction (RHEED), and the film with a thickness of 10 QL is obtained after ten periods of RHEED oscillation, as shown in Fig. 1(b). In the meantime, high-resolution scanning transmission electron microscopy (HRSTEM) is used to characterize the film structure and crystalline configuration [40]. Figure 1(c) highlights the highly-ordered hexagonal structure of the 10 QL $(Cr_{0.12}Bi_{0.26}Sb_{0.62})_2Te_3$ film with an atomically sharp interface on top of the GaAs substrate, and the uniform Cr distribution inside the host TI matrix is also confirmed by the energy dispersive X-ray (EDX) spectrum.

To investigate the chiral transport properties in the QAHE regime, we deliberately fabricate the Hall bar devices with dimensions of 2 mm × 1 mm (*i.e.,* ten times larger than that used in [30]), as shown in Fig. 1(a). In the diffusive transport region (T > 1 K), the 10 QL $(Cr_{0.12}Bi_{0.26}Sb_{0.62})_2Te_3$ film shows a typical semiconductor behavior, where the sample resistance monotonically increases as the sample temperature drops from 300 K to 1 K (Fig. 1(d)), indicating that the Fermi level is inside the bulk band gap [41]. Besides, the Curie temperature ($T_C$) is found to be around 30 K from the temperature-dependent magnetization under the field-cooled condition, as shown in Inset of Fig. 1(d) (Supplementary Information S1).

Figure 2 shows the magneto-transport results of the 10 QL $(Cr_{0.12}Bi_{0.26}Sb_{0.62})_2Te_3$ film. When $T < T_C$, the Hall resistance $R_{xy} = R_{14,62}$ in Fig. 2(b) develops a square-shaped hysteresis loop, indicating the robust FM order with an out-of-plane magnetic anisotropy, and the butterfly-shaped double-split longitudinal resistance $R_{xx} = R_{14,65}$ is also observed in Fig. 2(c) [28,29]. Our 10 QL $(Cr_{0.12}Bi_{0.26}Sb_{0.62})_2Te_3$ film reaches the QAHE regime when the sample temperature falls below 85 mK. As demonstrated in Fig. 2(b) and 2(c), the $R_{xy}$ reaches the quantized value of $h/e^2$ (25.8 k$\Omega$) at $B = 0$ T while $R_{xx}$ is nearly vanished. Meanwhile, it is important to highlight that the QAHE is also realized in the 6 QL (2D hybridization thickness) $(Cr_{0.12}Bi_{0.26}Sb_{0.62})_2Te_3$ film with a similar phase transition temperature of 85 mK (Supplementary Information S3). Therefore, the thickness-dependent results provide strong evidence that the stability of the QAHE phase in magnetic TIs is maintained as the film thickness varies across the hybridization limit (whereas in the QHE regime, the formation of the precise Landau level quantization requires the electrons to be strictly confined in the 2D region [42]).

To quantitatively understand the chiral edge transport in our samples, we thus apply the Landauer-Büttiker formalism that [43]

$$I_i = \frac{e^2}{h} \sum_j (T_{ji} V_i - T_{ij} V_j) \qquad (1)$$

where $I_i$ is the current flowing from the $i^{th}$ contact into the sample, $V_i$ is the voltage on the $i^{th}$ contact, and $T_{ji}$ is the transmission probability from the $i^{th}$ to the $j^{th}$ contacts [44,45]. In our six-terminal Hall bar structure shown in Fig. 2(a), the voltage is applied between the $1^{st}$ and $4^{th}$ contacts ($V_1 = V$, $V_4 = 0$), and the other four contacts are used as the voltage probes ($I_2 = I_3 = I_5 = I_6 = 0$). In the QAHE regime, since the TRS is broken, electrons can only flow one-way along the edge channel with the conduction direction determined by the magnetization orientation [39]. Specifically, when the film is magnetized along +z direction (left panel of Fig. 2(a)), the non-zero transmission matrix elements for the QAHE state are $T_{61} = T_{56} = T_{45} = 1$ [11,39], and the corresponding voltage distributions are given by $V_6 = V_5 = V_1 = (h/e^2)I$ and $V_2 = V_3 = V_4 = 0$. On the other hand, when the magnetization reverses its direction (right panel of Fig. 2(a)), the edge current flows through the 2$^{nd}$ and 3$^{rd}$ contacts, thus making $V_2 = V_3 = V_1 = (h/e^2)I$ and $V_5 = V_6 = V_4 = 0$. Consequently, $R_{14,62} = (V_6 - V_2) / I$ is positive for the $M_Z > 0$ case, and change to negative sign if $M_Z < 0$. Simultaneously, except for the sharp magneto-resistance (MR) peaks at the coercivity fields (±0.12 T), the vanishing $R_{12,65}$ in the fully magnetized region is also anticipated from Eq. (1) since the presence of the dissipationless chiral edge states lead to zero voltage drop along the edge channel. Accordingly, the consistency between the scenario described by Eq. (1) and the experimental observations in Fig. 2 clearly reveals the chiral edge transport nature of QAHE.

Compared with the QSHE helical states which were only observed in the μm-scale devices [10-12], the observation of scale-invariant QAHE with perfect quantization on the macroscopic scale is significant. Based on Eq. (1), the chiral nature in the QAHE regime causes $(T_{i,i+1}, T_{i+1,i}) = (0, 1)$. In other words, once the magnetization is fixed, the backward conduction is always prohibited by the chirality, and hence the de-coherence process from the lateral contacts cannot lead to momentum and energy relaxation [11,39]. Together with the thickness-dependent results, we may conclude that when appropriate spin-orbit interaction and perpendicular FM exchange strength are present in a bulk insulating magnetic TI film where the Fermi level resides inside the surface gap [26,27,33], the QAHE resistance is always quantized to be $h/e^2$, regardless of the device dimensions and de-phasing process (Supplementary information S4).

Figure 3(a) shows the $R_{xx}$-$T$ and $R_{xy}$-$T$ results of the 10 QL $(Cr_{0.12}Bi_{0.26}Sb_{0.62})_2Te_3$ film at $B = 3$ T and 15 T in the low-temperature region ($T < 1$ K). Both the enhanced magnetization and the reduced thermal activations at lower temperatures help localize the bulk conduction channels, and thus drive the system from the regular diffusive transport regime (T > 1 K) towards the chiral edge conduction regime. As a result, $R_{xx}$ diminishes rapidly as the sample temperature drops, which is opposite to the $R_{xx} - T$ relation in the higher temperature region as shown in Fig. 1(d). Moreover, when the magnetic TI film reaches the QAHE state below 85 mK, we also observe a non-zero $R_{xx}$ similar to the previous 5 QL $Cr_{0.15}(Bi_{0.1}Sb_{0.9})_{1.85}Te_3$ film case [30]. It is noted that the underlying mechanisms of the non-zero longitudinal resistances in these two systems are quite different. In particular, it was reported that when a large external magnetic field (B > 10 T) was applied, the 5 QL film was driven into a perfect QHE regime, and $R_{xx}$ diminished almost to zero[30]. In contrast, it is apparent from Fig. 3(a) that the longitudinal resistance in our 10 QL $(Cr_{0.12}Bi_{0.26}Sb_{0.62})_2Te_3$ sample remains at 3 kΩ even when the applied magnetic field reaches 15 T. More importantly, unlike the bulk conduction case which has a typical parabolic MR relation [46], $R_{xx}$ at $T = 85$ mK exhibits little field dependence when the magnetic field is larger than 3 T (Fig. 3(b)). Consequently, it may be suggested that the non-zero $R_{xx}$ in the thicker 10 QL magnetic TI film is more likely associated with a unique dissipative edge conduction, whose origins cannot be simply attributed to either the variable range hopping (VRH) [30] or the gapless quasi-helical edge states [39] as proposed for the 5 QL magnetic TI film.

To further confirm the presence of the dissipative edge conduction in addition to the QAHE state, we perform the non-local measurements on the six-terminal Hall bar device under different magnetic fields in Fig. 4 [39]. Two different non-local configurations are investigated: in case **A**, the current is passed through contacts 1 (source) and 2 (drain) while the non-local resistance between contacts 5 and 4 ($R_{12,54}$) is measured (top inset of Fig. 4(a)); in case **B**, a quasi H-bar geometry is adopted such that contacts 2 and 6 are designated as the source/drain pads while contacts 3 and 5 are used as the voltage probes ($R_{26,35}$, top inset of Fig. 4(b)) [39]. In the QAHE regime (T < 85 mK), it can be clearly seen that both $R_{12,54}$ and $R_{26,35}$ display square-shaped hysteresis windows with $H_C$ = 0.12 T, but their polarities are opposite. In other words, when B < -0.2 T, $R_{12,54}$ reaches the high-resistance state of 15 Ω while $R_{26,35}$ is at the low-resistance state close to zero. Here, we point out that the non-local resistances can be understood from the chirality of QAHE. In the inset of the bottom Fig. 4(a), for example, we show that when the film is magnetized along +z direction, the chirality forces the QAHE dissipationless current flow from contact 1 to contact 2 through the 1→6→5→4→3→2 contacts successively, and in turns "shorts" the contacts so that $V_6 = V_5 = V_4 = V_3 \sim V_1 = V$ [43]. As a result, the voltage drop between these contacts is negligible, and $R_{12,54}$ is driven into the low-resistance state (2 Ω). On the other hand, when the magnetization is reversed (-z direction), the 1$^{st}$ and 2$^{nd}$ contacts are directly connected through the upper edge (lower left panel of Fig. 4(a)), and the voltage probes from $V_3$ to $V_6$ are now away from the dissipationless QAHE channel. Consequently, the non-local signal only relates to the voltage drop caused by the dissipative edge channel, which gives rise to a larger value of $R_{12,54}$. The same transport principle can also be applied to the quasi H-bar non-local configuration (case **B**), and the illustrations of field-dependent conduction paths are consistent with the measured $R_{26,35}$ results, as shown in the bottom panels of Fig. 4(b) (Supplementary information S5). It is noted that, in contrast to the QAHE regime (T < 85 mK), both $R_{12,54}$ and $R_{26,35}$ are dominated by the larger bulk conduction component at a higher temperature of 4.7 K (*i.e.,* the non-local resistances are more than 10 times larger than those probed at 85 mK). In such diffusive transport regime, the square-shaped hysteresis non-local signals are replaced by the ordinary parabolic MR backgrounds, and the polarity differences between $R_{12,54}$ and $R_{26,35}$ also disappear. To summarize, both the field-independent $R_{xx}$ shown in Fig. 3(b) and the non-local resistances displayed in Fig. 4 confirm the coexistence of QAHE chiral edge channel and the additional dissipative edge conduction in the thick 10 QL Cr-doped TI sample.

In a uniformly doped 3D magnetic TI, the out-of-plane magnetization opens gaps on the top and bottom surfaces, and changes each surface into a non-zero $C_1$ phase with a half-quantized Hall conductance ($e^2/2h$) [7,21,31,32]. It is noted that the gapped Dirac fermions on these two surfaces have opposite masses and topological characters because of their opposite normal vector directions [7]. Under such circumstances, the side surface acts as the domain wall between the two chiral edge bands, and the total quantum Hall conductance of $e^2/h$ is observed since both the top and bottom surface edges are contacted by the same electrode in our experiment [7,31,32]. Meanwhile, the gapless Dirac point on the side surfaces is shifted away from the symmetric point ($\Gamma$) due to the perpendicular magnetization (Supplementary information S6). Therefore, the backscattering is not suppressed anymore, and the side surface conduction thus becomes dissipative [31], which is intimately related to the dissipative edge channel unveiled in Fig. 3 and 4. Nevertheless, further investigations are needed to elucidate the intrinsic mechanisms of the dissipative edge states in the 3D magnetic TI system. Systematic thickness-dependent experiments as well as relevant theoretical modeling are required to address this issue in more detail.

In conclusion, our results demonstrate QAHE in the magnetic TI material with the thickness beyond the 2D hybridization limit. Both the scale-invariant dissipationless Hall conductance and two-state non-local resistances not only reflect the chiral transport character of the QAHE state, but also reveal the distinctions between QAHE and the other two quantum phases (*i.e.,* QHE and QSHE). Moreover, a dissipative channel is observed for the thick 10 QL magnetic TI sample and its origin is different from the 2D hybridization thickness case. The scale-invariant feature of QAHE may motivate the exploration of new QAHE phases and may provide novel chiral interconnects with higher quantized conduction channels.

**Note added**. During the preparation of the manuscript, we are aware of a related work by Checkelsky *et al.* [47] that reports the observation of QAHE in a 8 QL Cr-doped magnetic TI.

*Acknowledgements* --- We thank Dr. J. Wang, Prof. X.L. Qi, Prof. D. Goldhabor-Gordon, Prof. Y. Tserkovnyak, and Prof. Y.G. Yao for helpful discussions. We are grateful to the support from the DARPA Meso program under contract No.N66001-12-1-4034 and N66001-11-1-4105. We also acknowledge the support from the FAME Center, one of six centers of STARnet, a Semiconductor Research Corporation program sponsored by MARCO and DARPA. K.L. W acknowledges the support of the Raytheon endorsement. X. K and M. L acknowledge partial support from the Qualcomm Innovation Fellowship. W.L. L. acknowledges funding support from the Academia Sinica 2012 career development award in Taiwan. Y. W acknowledges support from Natural Science Foundation of China (11174244 and 51390474) and Zhejiang Provincial Natural Science Foundation (LR12A04002) and National 973 project of China (2013CB934600).

**Figures and Captions**

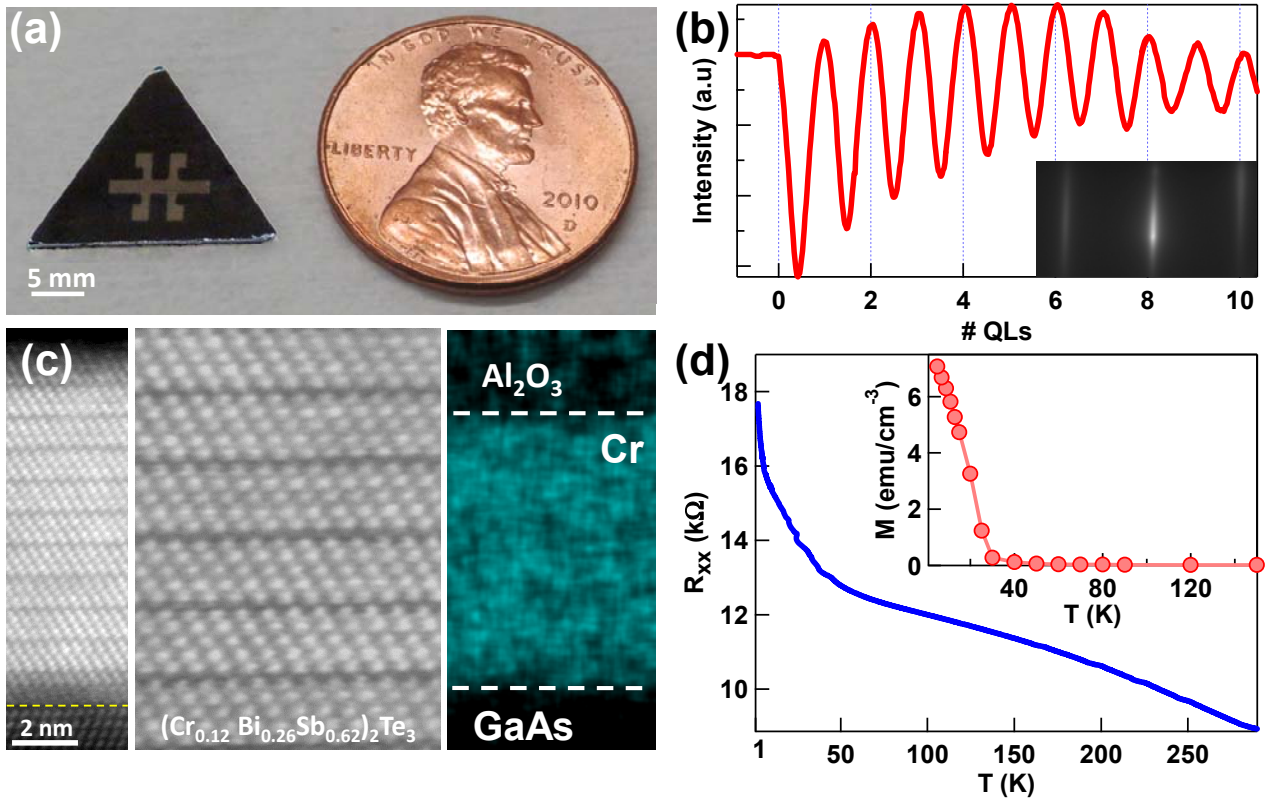

**FIG. 1 (color online). Cr-doped $(Bi_xSb_{1-x})_2Te_3$ film structure and properties. (a)** The image of the Hall bar-structure with the dimension of 2 mm × 1 mm. **(b)** RHEED oscillation, showing that the grown Cr-doped $(Bi_xSb_{1-x})_2Te_3$ film has a thickness of 10 QL. Inset: RHEED pattern of the as-grown film. **(c)** Cross-sectional HRSTEM image, illustrating the crystalline structure of the 10 QL $(Cr_{0.12}Bi_{0.26}Sb_{0.62})_2Te_3$ film. The EDX mapping confirms that the Cr dopants distribute uniformly inside the TI layer. **(d)** Temperature-dependent resistance as temperature drops from 300 K to 1 K. Inset: Magnetic moment under field-cooled condition. The applied field is 10 mT and the Curie temperature is estimated to be $T_C$ = 30 K.

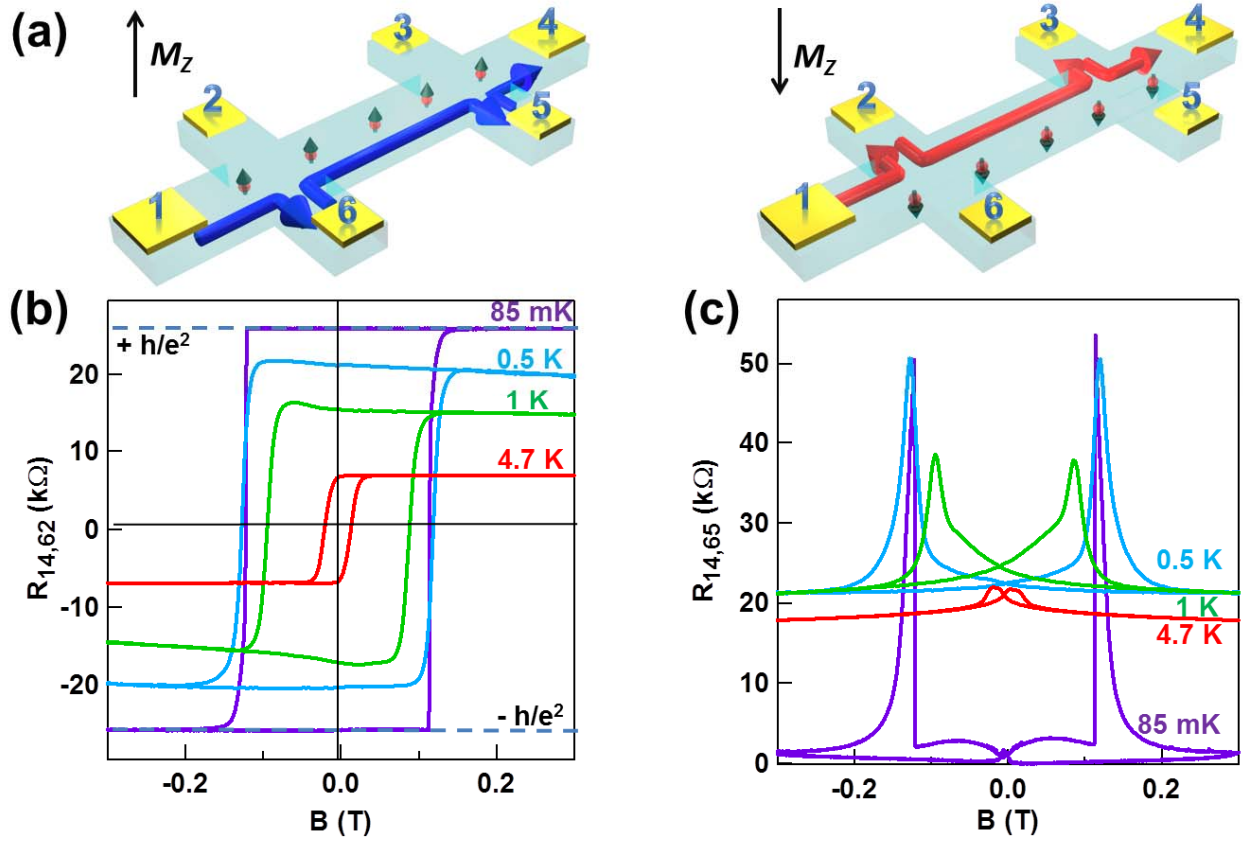

**FIG. 2 (color online). QAHE in the 10 QL $(Cr_{0.12}Bi_{0.26}Sb_{0.62})_2Te_3$ thin film. (a)** Schematics of the chiral edge conduction in the QAHE regime. The current flows from the 1st contact to the 4th contact, and the magnetization of the Cr-doped TI film is along the z-direction. **(b)** Hysteresis $R_{xy}$–$B$ curves at different temperatures. For T < 85 mK, $R_{xy}$ attains the quantized value of $h/e^2$. **(c)** Butterfly-shaped $R_{xx}$–$B$ curves. In the QAHE regime, $R_{xx}$ nearly vanishes at low fields.

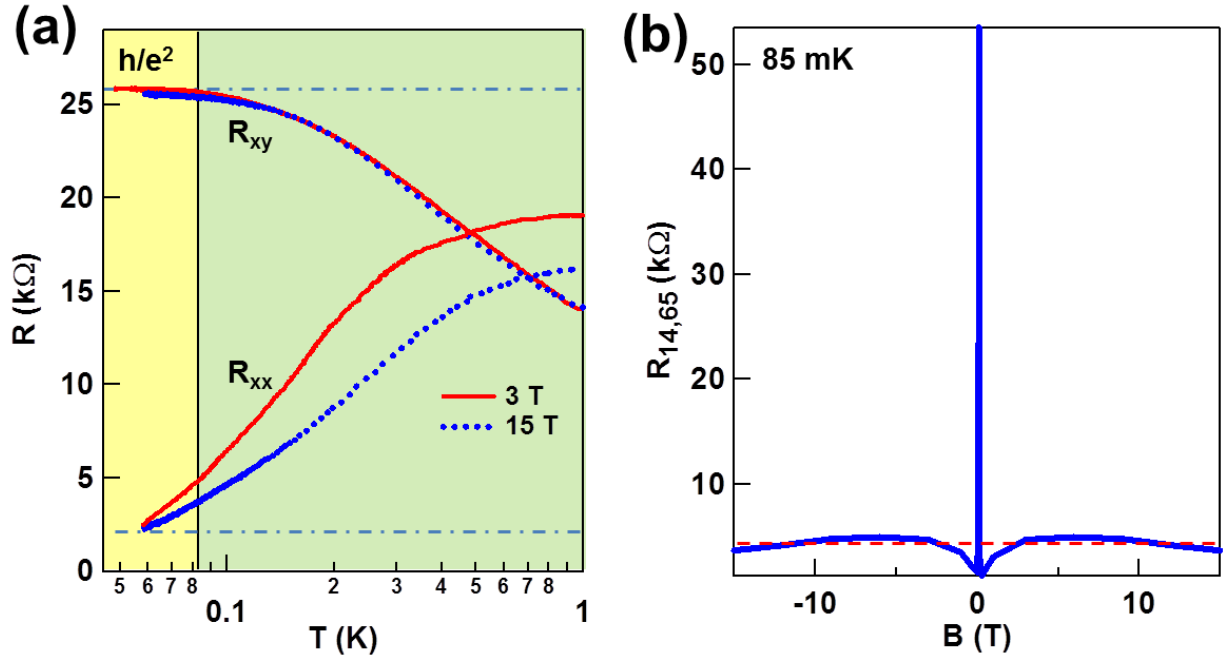

**FIG. 3 (color online). Non-zero longitudinal resistance in the QAHE regime. (a)** Temperature-dependent $R_{xx}$ and $R_{xy}$ of the 10 QL $(Cr_{0.12}Bi_{0.26}Sb_{0.62})_2Te_3$ film at $B = 3$ T and 15 T in the low-temperature region. **(b)** Magnetic field dependence of $R_{xx}$ at 85 mK. $R_{xx}$ in our 10 QL magnetic TI sample shows little field dependence when B > 3 T.

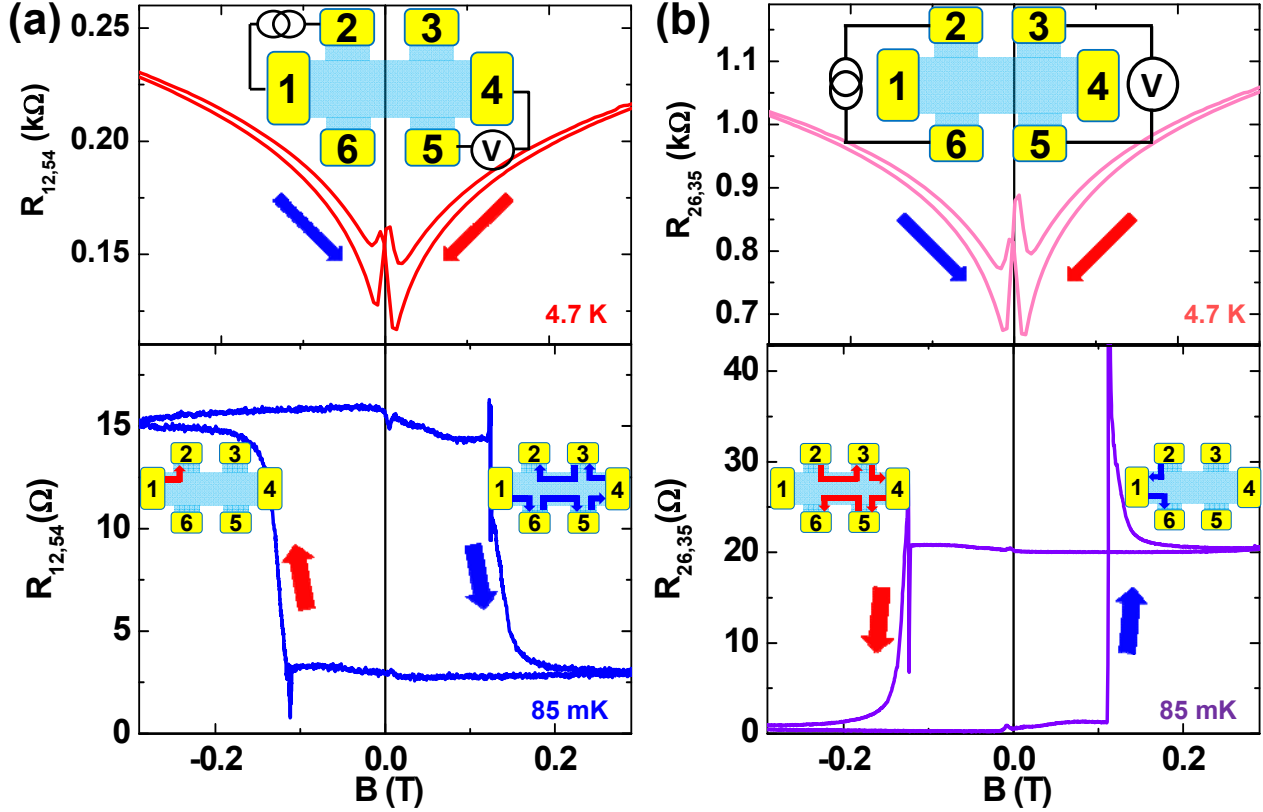

**FIG. 4 (color online). Non-local measurements of the six-terminal Hall bar device. (a)** Case **A**: the current is applied through 1st to 2nd contact and the non-local voltages are measured between 4th and 5th contacts at $T$ = 85 mK and 4.7 K. **(b)** Case **B**: the current is applied through 2nd to 6th contact and the non-local voltages are measured between 3rd and 5th contacts with different magnetizations. Inset: Illustrations of the QAHE channel under different magnetizations. The red solid arrow indicates that the magnetic field is swept from positive (+z) to negative (-z), whereas the blue solid arrow corresponds to the opposite magnetic field sweeping direction.